\documentclass[12pt,a4paper]{article}
\usepackage[a4paper,margin=2.5cm]{geometry}
\usepackage{setspace} 
\usepackage{authblk}
\usepackage[square,sort,comma,numbers]{natbib}
\usepackage{booktabs}
\usepackage{hyperref}
\usepackage{amsmath} 
\usepackage{amssymb}
\usepackage{graphicx}
\usepackage{float}

\date{}

\usepackage{chngcntr}
\counterwithout{equation}{section}

\title{\textbf{Astrophysical Signatures of Fermionic Dark Matter}}
\author[1]{Vinit D Tyagi}
\author[1]{Suhas S S}
\author[1]{Arun Kenath\textsuperscript{*}}
\affil[1]{\small{Department of Physics and Electronics, Christ University, Bengaluru, India}}

\begin{document}

\thispagestyle{empty}
  \maketitle
  \begin{center}
    \textsuperscript{*}Corresponding Author: kenath.arun@christuniversity.in
  \end{center}
  \begin{abstract}
   Fermionic dark matter particles remain one of the most compelling candidates for the dark matter content of the Universe, yet no positive results have been obtained from direct detection experiments. In this work, we investigate the possibility that such particles form compact gravitationally bound objects supported by degeneracy pressure. Through an extensive review of microlensing surveys, we derive constraints on the masses of these compact objects. We further analyze the accretion of baryonic matter onto these objects and evaluate their thermal and radiative properties. The estimated burst emission is found to be well below the energies associated with the Galactic Center GeV excess, suggesting that these compact objects are unlikely to be the source of the observed signal. Our analysis suggests that admixed dark matter and baryonic matter objects could potentially account for a fraction of the presently unobserved baryonic matter, thereby providing a possible explanation for a fraction of the missing baryons in the Universe.

  \end{abstract}


\setcounter{page}{1}
\newpage
\section{Introduction}
Dark Matter (DM) is a form of non-luminous matter that interacts predominantly via gravity (and possibly via other very weak forces) but not electromagnetically. Its existence is inferred from multiple gravitational phenomena on astrophysical scales. Precision cosmological measurements show that ordinary Baryonic Matter (BM) constitutes only about 5\% of the universe’s energy density, whereas roughly 27\% is in DM \cite{aghanim2020planck} \cite{Arun_2017}. This unseen component dominates the mass of galaxies and clusters, yet its particle nature remains a major open question in cosmology and particle physics.\\
Multiple lines of indirect evidence point to DM. One of the earliest pieces of evidence was provided by Zwicky’s 1933 study of the Coma Cluster. In spiral galaxies, the orbital velocities of stars and gas remain approximately constant at large radii, rather than falling off as expected from the visible mass distribution. These flat rotation curves imply a massive, extended halo of unseen matter \cite{RubinDMevidence}. Ordinary BM alone cannot account for the inferred gravitational potential. The pattern of temperature fluctuations in the cosmic microwave background is sensitive to the total matter content. Precision data from WMAP and Planck \cite{aghanim2020planck} require DM to fit the heights and positions of the acoustic peaks and baryons alone cannot reproduce the observed CMB spectrum.\\
Another way we can infer DM is through Gravitational Lensing. Light from background sources is deflected by the gravitational fields of foreground galaxies and clusters. Lensing measurements consistently reveal more mass than is visible. For example, in the Bullet Cluster (1E0657-558) collision, observations show that the regions with the most mass determined through gravitational lensing do not align with the hot X-ray-emitting gas, which contains most of the BM. Instead, the mass is concentrated around the galaxies, which passed through each other during the collision largely unaffected. This clear separation between visible matter and gravitational mass provides strong evidence for the existence of DM \cite{bulletDM}.\\
The DM candidates can be initially broadly classified as Hot DM and Cold DM. Hot Dark Matter consists of relativistic particles (such as standard-model neutrinos). They free stream out of overdense regions, erasing structure on small scales and leading to top-down formation, which is incompatible with the observed early formation of galaxies. Whereas, Cold Dark Matter refers to particles that were non-relativistic at early times (e.g.,\ WIMPs or Axions). CDM leads to bottom-up structure formation, where small halos form first and merge into larger systems. This scenario is consistent with observations of galaxies, groups, and clusters. Warm Dark Matter represents the intermediate scenario between these two extremes. But the success of CDM models in matching both the CMB and the galaxy distribution strongly suggests that the dominant dark matter component must be cold and is the preferred model.\\
Prominent candidates include Weakly Interacting Massive Particles (WIMPs) and Axions, both motivated by physics beyond the Standard Model. Axions are very light ($\mu$ev to 1 eV scale) particles originally proposed to solve the strong-CP problem in QCD \cite{axionRef}.They are produced non-thermally in the early universe and act as Cold Dark Matter.\\
Weakly interacting massive particles (WIMPs) are prominent theoretical candidates for DM \cite{STEIGMAN1985375}, characterized by their substantial mass (ranging from 10 to 1000 GeV) and weak interactions with ordinary matter via the weak nuclear force and gravity. \\
There are many other probable candidates like Primordial Blackholes, Sterile Neutrinos, Fermi Balls and even alternate theories for DM like Modified Newtonian Dynamics (MOND) propose altering gravity at low accelerations which is successful in explaining some of the observations like the galactic rotation curves.\\
Beyond treating dark matter as individual particles, another possibility is that it may form compact, self-gravitating objects under suitable conditions. If the dark matter constituent particles are fermionic in nature, they obey the Pauli exclusion principle, which generates degeneracy pressure at high densities \cite{Ruffini1969,Narain2006}. This pressure can counterbalance gravitational collapse at high densities, similar to what is observed in white dwarfs and neutron stars. Therefore, fermionic dark matter can potentially form stable compact objects with characteristic masses determined by the particle mass. In this work, we explore such configurations and constraint their mass using microlensing observations. \\
There has been both direct and indirect attempts to search for these DM particles. Direct detection aims to observe dark matter particles scattering off atomic nuclei in ultra sensitive underground detectors by measuring the resulting nuclear recoil. Indirect detection searches for the byproducts of dark matter annihilation or decay such as gamma rays, neutrinos, or antimatter in cosmic ray or astrophysical observations.

\subsection{Direct Detection of WIMPs}
Weakly Interacting Massive Particles (WIMPs) interact through the weak force and the gravitational force. They have a wide range of masses and due to these elusive properties, numerous direct detection experiments have been conducted to observe WIMPs through their weak nuclear interactions; however, none have yielded positive results to date \cite{undagoitia2015dark}.\\
One method for detecting WIMPs is the direct‐detection approach. These experiments aim to measure the tiny energy transferred when a halo WIMP scatters elastically off a target nucleus.  In the center‐of‐mass frame, the nuclear recoil energy is given by \cite{DiGangi2021},
\begin{equation}
    E_{R} = \frac{|\mathbf{q}|^2}{2 m_{N}} =
\frac{\mu^2 v^2}{m_{N}} \bigl(1 - \cos\theta_{R}\bigr) ,
\end{equation}
where $m_N$ is the target‐nucleus mass, $\mu = m_\chi m_N/(m_\chi + m_N)$ is the reduced mass, $v$ is the WIMP speed in the laboratory frame, and $\theta_{R}$ is the scattering angle.  The minimum velocity required to induce a recoil of energy $E_R$ follows by setting $\cos\theta_{R}=-1$:
\begin{equation}
    v_{\min} =\sqrt{\frac{m_{N} E_{R}}{2 \mu^{2}}} = \frac{|\mathbf{q}|}{2 \mu} 
\end{equation}
Now, the differential event rate for simplified WIMPs interaction is given by,
\begin{equation}
    \frac{dR}{dE_R}= \frac{R_o}{E_o r}e^{\frac{-E_R}{E_o r}}
\end{equation}
where, LHS is the event rate, $E_o$ is the most probable event energy, $R_o$ is the total event rate and r is the kinematic factor,
\begin{equation}
    r= \frac{4m_{\chi} m_N}{(m_{\chi}+ m_N)^2} 
\end{equation}
Under the assumption of a local DM density $\rho_{0}$ and a velocity distribution $f(v)$, the differential recoil rate per unit detector mass can be written as \cite{cerdeno2010direct, cerdeno2013nuclear}
\begin{equation}
    \frac{dR}{dE_{R}} = \frac{\rho_{0}}{m_{\chi} m_{N}} \int_{v_{\min}}^{\infty} v f(v)  \frac{d\sigma_{\chi N}}{dE_{R}}  dv 
\end{equation}
Here $m_{\chi}$ is the WIMP mass and $ \frac{d\sigma_{\chi N}}{dE_{R}}$ is the differential scattering cross section.  
Finally, the cross section separates into spin‐independent (SI) and spin‐dependent (SD) components:
\begin{equation}
    \frac{d\sigma_{\chi N}}{dE_{R}} = \biggl[\frac{d\sigma}{dE_{R}}\biggr]_{\!\mathrm{SI}} + \biggl[\frac{d\sigma}{dE_{R}}\biggr]_{\!\mathrm{SD}} 
\end{equation}
The SI term typically scales as $A^{2}$, with $A$ the nuclear mass number, while the SD term depends on the nuclear spin content.  In practice, nuclear form factors, detector efficiencies, and energy‐resolution effects must be included when comparing these theoretical rates to experimental data.

\subsubsection{Direct Detection of DM with Xenon}
Liquid Xenon is used as a medium for the detection as it is radiopure, i.e., it has many stable isotopes, has a high triple point and is an excellent scintillator \cite{DiGangi2021}. \\
From the XENON10 TPC detector built in 2005 to the XENONnT detector built in 2020, this experiment has made leaps of progress in terms of sensitivity and reduction of electronic recoil (ER). There are other experiments which use Xenon as the medium like Large Underground Xenon (LUX) experiment in US or DAMA experiment which uses thallium doped Sodium Iodide.\\
The effectiveness of these experiments is reflected in the stringent limits they have placed on dark matter interactions.
At 90\% confidence level, spin‐independent WIMP–nucleon cross sections are constrained by XENONnT to $\sigma_{SI}<2.58\times10^{-47} \mathrm{cm}^2$ at $m_\chi=28 \mathrm{GeV}/c^2$ \cite{aprile2023first} and by XENON1T to $\sigma_{SI}<4.1\times10^{-47} \mathrm{cm}^2$ at $m_\chi=30 \mathrm{GeV}/c^2$ \cite{Aprile_2018}. In the spin‐dependent channel, XENON100 excludes inelastic WIMP–nucleon interactions above the $\sigma_{SD}^{\rm inel}<3.3\times10^{-38} \mathrm{cm}^2$ at $m_\chi=100 \mathrm{GeV}/c^2$ \cite{aprile2017search}, while DAMA has reported an estimated mass of the DM particles would range between 10 to 15 GeV or between 60 to 100 GeV depending on the actual nucleus involved in the scattering process (sodium or iodine, respectively) \cite{roszkowski2018wimp}.

\subsection{Gravitational Microlensing}
Gravitational Lensing is a phenomenon where light from a distant source, such as a star or galaxy, is bent as it passes near a massive object \cite{wambsganss1998gravitational} . Gravitational lensing happens because mass distorts the fabric of
space-time. This effect is most pronounced around extremely massive
objects, such as black holes and entire galaxies. However, smaller objects, such as stars and planets, also produce a measurable distortion, known as microlensing \cite{paczynski1986gravitational}.\\ 
Microlensing is an incredibly powerful tool in astrophysics that helps us gain new perspectives into the nature of objects which are difficult to see. Some of the uses of microlensing include its usage in the study of DM and galaxies, and for the discovery of planets outside our solar system. Microlensing is expected to play an important role in future discoveries in astrophysics. The mass of compact dark matter objects can be estimated using microlensing events, and this information can indirectly help us estimate the mass of DM particles.
\subsubsection{Optical Depth}
The optical depth for microlensing $\tau$, is a measure of the probability that a given line of sight to a star will intersect the Einstein radius of a compact lensing object. The optical depth up to a given source distance $D_S$ represents the probability that the line of sight to a target source intersects the Einstein disk of a lensing object at any given moment, resulting in a magnification of A$>$1.34\cite{Tisserand_2007} . assuming that the distribution of deflector masses is given by the density function \( \rho(D_L) \) and the normalized mass function \( \frac{dn_L(D_L, M)}{dM} \), the probability can be expressed as \cite{moniez2010microlensing}:

\begin{equation}
    \tau(D_S) = \int_0^{D_S} \int_0^{\infty} \pi \theta_E^2 \times 
\frac{\rho(D_L) D_L^2}{M} \frac{dn_L(D_L, M)}{dM}   dM   dD_L
\end{equation}
where \( \theta_E = \frac{R_E}{D_L} \) is the angular Einstein radius of a lens with mass \( M \) located at \( D_L \).  
The second term of the integral represents the differential number of these lenses per unit mass and per unit solid angle.
also, solid angle of Einstein ring is directly proportional to lens's mass M, optical depth is found to be independent of mass function.

\begin{equation}
    \tau(D_S) = \frac{4\pi G D_S^2}{c^2} \int_0^1 x (1 - x) \rho(x)   dx
\end{equation}
where $\rho(x)$
 is mass density of lens located at distance of $xD_s$.
The microlensing optical depth \( \tau \) for a given field of view can be estimated from observational data. For example, the MACHO (Massive Compact Halo Object) collaboration found an optical depth towards the Large Magellanic Cloud (LMC) of about \( \tau_{\text{obs}} \approx 1.2 \times 10^{-7} \) \cite{alcock}.
\subsubsection{Event Rate}
The event rate in microlensing describes how frequently microlensing events occur within a given field of view. It depends on the number of potential lensing objects, their velocities, and the optical depth. The optical depth measured from microlensing events observed in a population of $N_{\text{obs}}$ stars, monitored over a duration of $T_{\text{obs}}$, is usually given by \cite{moniez2010microlensing}:
\begin{equation}
    \tau = \frac{1}{N_{\text{obs}} \Delta T_{\text{obs}}} \times \frac{\pi}{2} \sum_{\text{events}} \frac{t_E}{\epsilon(t_E)}
\end{equation}
where $\epsilon(t_E)$ is the average detection efficiency of microlensing events with a time scale $t_E$.
The total measured event rate, adjusted for detection efficiency, is usually given by:
\begin{equation}
    R = \frac{1}{N_{\text{obs}} \Delta T_{\text{obs}}} \sum_{\text{events}} \frac{1}{\epsilon(t_E)}.
\end{equation}
This event rate and duration of events are used to constrain the mass of lens.

\section{Constraining the Mass of DM Object}
\begin{figure}[!h]
\centering
\includegraphics[width=12.5cm]{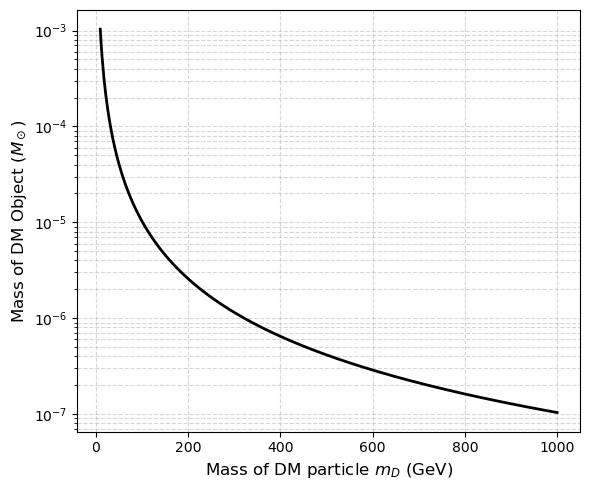}
\caption[Plot between Mass of DM object and Mass of DM particle]{Plot between Mass of DM object and Mass of DM particle}
\label{fig:DMObjectvsDMmass}
\end{figure}

The characteristic mass scale of these fermionic dark matter compact objects can be understood using an energy balance argument, In such systems, gravity acts to compress the object while the Pauli exclusion principle gives rise to quantum kinetic energy arising from the confinement of fermions within a finite volume. As the object contracts, the momenta of the fermions increase, which in turn enhances the quantum pressure. This pressure opposes further gravitational collapse and plays a key role in stabilizing the system. By balancing the gravitational energy of the system with this quantum kinetic energy, this leads to a characteristic maximum mass for stability. the scaling takes the form  $M \sim \frac{M_{\mathrm{Pl}}^3}{m_\chi^2}$,
where $M_{\mathrm{Pl}}$ denotes the Planck mass and $m_\chi$ represents the mass of the fermionic dark matter particle. This relation is analogous to the Chandrasekhar mass limit for white dwarfs. Such scaling relations have been discussed in the context of fermionic dark matter compact objects \cite{Ruffini1969,Narain2006}. \\
For our objects  which is made up of fermionic DM particles, Chandrasekhar mass is given by,
\begin{equation}
    M_{D(CH)} = (\frac{hc}{2\pi G})^{\frac{3}{2}}\frac{1}{m_D^2} 
\end{equation}
where $m_D$ is mass of dark matter particle. We have considered a range of fermionic dark matter particle masses in the range of 10 GeV to 1000 GeV, this range spans representative GeV scale dark matter candidates frequently investigated in dark matter phenomenology. Using eq. (11) the corresponding compact object masses are found to be $3.25 \times 10^{31}g$, $3.25 \times 10^{29}g$ and $3.25 \times 10^{27}g$ for 10 GeV, 100 GeV, and 1000 GeV respectively. The mass of the object decreases inversely with the square of the particle mass, as shown in Fig.~\ref{fig:DMObjectvsDMmass}. We will constrain this and check which of these objects are feasible.
\\
\noindent A thorough review of the literature demonstrates that the candidate objects under consideration cannot be classified as Massive Astrophysical Compact Halo Objects (MACHOs) \cite{Green_2016} \cite{Basak_2022}. Additionally, observational data is further confirming the existence of unbound objects that are present within the range of detection of our survey \cite{szymański2011opticalgravitationallensingexperiment}. Furthermore, theoretical calculations show that the mass of such objects can be as great as the mass of asteroids or planets \cite{primordial}.

\section{Microlensing Signatures}

Gravitational microlensing occurs when a compact lensing object passes close to the line of sight to a more distant source, producing a transient magnification of the source’s light.  The Einstein radius, defined as the characteristic scale of this effect, is given by:
\begin{equation}
    R_E = \sqrt{\frac{4 G M}{c^2} \frac{D_L \times (D_{S} - D_L)}{D_S}}
\end{equation}
where \(G\) is the gravitational constant, \(c\) the speed of light, \(M\) the mass of the lens, and \(D_L\) and \(D_S\) are the observer–lens and observer–source distances, respectively \cite{paczynski1986gravitational}.  The observable duration of a microlensing event is the time required for the lens to traverse this radius at a transverse velocity \(v\):
\begin{equation}
    t= \frac{R_E}{v}
\end{equation}
To constrain the lens mass using microlensing observables, we compute the differential event rate.  For a single source star and unit exposure, the rate per unit lens position \(x = D_L/D_S\) and event timescale \(t_E\) is given by \cite{Croon_2020}:
\begin{equation}
    \frac{d^2 \Gamma}{dxdt_E} = \epsilon (t_E) \frac{2D_s}{v_0^2 M }f_{DM} \rho_{DM}(x) v_E^4 (x) e^{\frac{-v_E^2 (x)}{v_0^2}}
\end{equation}
Here \(v_0 = 220\ \mathrm{km/s}\) denotes the characteristic dark‐matter circular speed in the Galaxy, \(\epsilon(t_E)\) the detection efficiency at timescale \(t_E\), \(f_{DM}\) the fraction of dark matter in compact lenses, and \(\rho_{DM}(x)\) the halo density profile.  Integrating over \(x\) and \(t_E\) yields the total expected number of events:
\begin{equation}
    N_{events} = N_{*}T_{obs} \int^1_0 dx \int^{t_{E,max}}_{t_{E,min}} dt_E \frac{d^2\Gamma}{dxdt_E} 
\end{equation}
where \(N_{*}\) is the number of monitored source stars, \(T_{obs}\) the total observation time, and \(t_{E,\min}\) and \(t_{E,\max}\) define the survey’s sensitivity window.
An alternative estimate uses the optical depth, defined by \cite{Tisserand_2007},
\begin{equation}
    \tau = D_s \int^1_0 dx \frac{f_{DM} \rho_{DM}(x)}{M} \pi r^2_E (x) 
\end{equation}
from which the expected number of events in the EROS survey is,
\begin{equation}
    N_{events}^{EROS} = N_{*} T_{obs} \frac{2 \epsilon(t_E)}{\pi \langle t_E \rangle } \tau 
\end{equation}
For a lens mass of \(1.63 \times 10^{-2} M_{\odot}\), the number of events predicted is 96.5 events, while the optical‐depth method yields 179.02 events.  The discrepancy arises because the former integrates over the full timescale distribution, whereas the latter employs the mean timescale. We also evaluated lens masses of $1.63\times10^{-6}\,M_\odot$ and $1.63\times10^{-4}\,M_\odot$. For a fixed halo fraction, these lower masses produce substantially shorter Einstein crossing times and consequently predict significantly larger numbers of microlensing events than the $1.63\times10^{-2}\, M_\odot$ case. Since such large event rates are not observed by the EROS-2 survey~\cite{Tisserand_2007}, these lower masses are subject to stronger microlensing constraints. Accordingly, among the masses considered in the present work, we adopt $1.63\times10^{-2}\, M_\odot$ as the benchmark lens mass for the subsequent analysis.

\section{Evolution and Energetics of DM objects}

Now that we have determined the feasible mass of the DM object to be $1.63\times10^{-2} M_{\odot}$, corresponding to a fermionic dark matter particle of 10 GeV, we examine its structural evolution and mass-accretion process following \cite{KIREN20212050}.

\subsection{Radius and Total Mass Accreted}

By knowing the object's mass, its radius can be estimated \cite{Sivaram_2011}:
\begin{equation}
    R = \frac{92 \times \hbar^2}{M^{1/3}  G  m_d^{8/3}}
    \label{radius}
\end{equation}
For $M = 1.63\times10^{-2} M_{\odot}$ and $m_d = 10 \mathrm{GeV}$, this yields $R = 2.2\times10^5 $cm.\\
The Bondi accretion rate is given by \cite{1952MNRAS.112..195B},
\begin{equation}
    dM = 4\pi R(t)^2 \rho_{DM}(t) v dt
    \label{bondi}
\end{equation}
with the time‐dependent background density \cite{rebecca2020dark},
\begin{equation}
    \rho(t) = \rho_o(t_o) \bigl(\tfrac{t_o}{t}\bigr)^2
    \label{density1}
\end{equation}
where $\rho_o$ is the density at early epoch $t_o$.
Substituting eqns ~\eqref{density1} and ~\eqref{radius} in ~\eqref{bondi} and integrating we get,
\begin{equation}
    M^{\frac{5}{3}} = M_o^{\frac{5}{3}} + 4 \pi v \rho_o (\frac{92 \times \hbar^2}{G \times m_d^{\frac{8}{3}}})^2 t_o 
\end{equation}
The first term in the RHS of the equation is the initial mass of the object, in this case $1.63 \times 10^{-2} M_{\odot} $, which we can ignore for now since we are looking at how much Hydrogen and Helium will be accreted. Therefore, the estimated masses of the accreted hydrogen and helium components are $M_H = 5.692\times10^{22} $g and $M_{He} = 1.776\times10^{23} $g.

\subsection{Thermal Structure, Density Limits, and Energetics}
With the beginning of accretion of hydrogen and helium onto the DM object, the gravitational energy gets converted to thermal energy, and the temperature of the accreting matter increases. An estimate of the temperature can be made by setting the gravitational potential energy equal to thermal energy per particle.
\begin{equation}
    T = \frac{G M m}{R k_B}
\end{equation}
where $k_B$ is the Boltzmann constant. Substituting the relevant parameters into Eq (22) yields a temperature of $T_H = 1.192\times10^{11} $K and $T_{He} = 4.770\times10^{11} $K.\\
During accretion, an estimate of the maximum density of the deposited layers can be obtained by balancing gravitational and radiation pressures.  The gravitational pressure is given by,
\begin{equation}
    P = \frac{G M_{DM} M_{acc}}{R^4} ,
\end{equation}
which increases as the layer density grows. simultaneously, the rising temperature from continued mass infall elevates the radiation pressure,
\begin{equation}
    P_{rad} = \frac{\sigma T^4}{c} 
\end{equation}
Equating $P = P_{\rm rad}$ and using $M_{acc} = 4\pi R^2 h \rho$, yields the maximum attainable density,
\begin{equation}
    \rho = \frac{\sigma T^4 R^2}{4\pi G M_{DM} h c} .
\end{equation}
The corresponding maximum densities associated with the hydrogen and helium components accreted are estimated to be $\rho_H = 3.20 \times 10^{13} \mathrm{g/cc}$ and $\rho_{He} = 1.26 \times 10^{15} \mathrm{g/cc} .$\\
Note that these density estimates are obtained from a simplified pressure-balance model and represent characteristic densities associated with the accreted hydrogen and helium components. At such extreme densities, the physical state of the material is expected to differ significantly from that of ordinary atomic hydrogen and helium. \\
To estimate the thickness of layers we use pressure scale height
\begin{equation}
    h= \frac{R_g T}{g} = \frac{R_g T R^2}{G M}
\end{equation}
Using Eq. (26), the corresponding scale heights are found to be 20.8 cm and 138 cm for the hydrogen and helium components, respectively.
These scale heights are much smaller than the radius of the compact object, indicating that the accreted material remains confined to thin surface layers.\\
Now, the mass accreted on these layers with thickness '$h$' will be 
\begin{equation}
    M_{acc} = 4\pi R^2 h \rho
\end{equation}
substituting the corresponding values into Eq (27) yields $M_{acc(H)} = 4.12\times10^{26} $g and $M_{acc(He)} = 1.05\times10^{29} $g.
\subsection{Eddington Luminosity and Radiative Energetics}

The Eddington luminosity $L$ is the maximum steady radiative output for an accreting object of mass $M$ and is given by,
\begin{equation}
    L = \frac{4\pi G M m_p c}{\sigma_T}
\end{equation}
Where $\sigma_T =6.65 \times 10^{-25}$ $cm^2$ is the Thomson cross section. For $M = 10^{-2} M_{\odot}$ this evaluates to $L = 2.052 \times 10^{36} \mathrm{erg/s}$. Since the fusion of $1 $g of hydrogen to helium releases $3\times10^{18} \mathrm{erg}$, sustaining $10^{36} \mathrm{erg/s}$ would require the processing of $10^{18} $g of hydrogen (and equivalently helium). Accretion beyond this limit may be inhibited as radiation pressure approaches the gravitational force.\\
The maximum energy released by accreting layers of hydrogen and helium is \\ $E_H      = M_{acc(H)}  \times 3\times10^{18} = 1.236\times10^{45} \mathrm{erg}$ and $E_{He}   = M_{acc(He)} \times 10^{18} = 1.05\times10^{47} \mathrm{erg}$.
We have characterized the structural evolution and energy output of objects composed of fermionic dark matter particles; the principal results are presented in the Table~\ref{tab:parameters}.
\renewcommand{\arraystretch}{1.5}
\begin{table}[h]
    \centering

    \begin{tabular}{|c|c|c|}
        \hline
        Parameters & Hydrogen & Helium \\
        \hline
        Total Mass accreted & $5.692 \times 10^{22}$g &$1.776 \times 10^{23}$ g\\
        \hline
        Temperature of layer & 1.192 $\times 10^{11}$K & 4.770 $\times 10^{11}K$  \\
        \hline
        Thickness accreted & 20.8 cm & 138 cm \\
        \hline
        Maximum Density & 3.20 $\times 10^{13} g/cc $ & 1.26 $\times 10^{15} g/cc $  \\
        \hline
        Mass accreted for thickness 'h' & 4.12 $\times 10^{26}g $ & 1.05 $\times 10^{29}g$ \\
        \hline
        Energy Released & $1.236 \times 10^{45} erg $ & $1.05 \times 10^{47}erg $ \\
        \hline
    \end{tabular}
    \caption{Parameters of the DM object}
    \label{tab:parameters}
\end{table}

\section{Galactic Center GeV Excess}

\begin{figure}[H]
    \centering
    \includegraphics[width=0.6\linewidth]{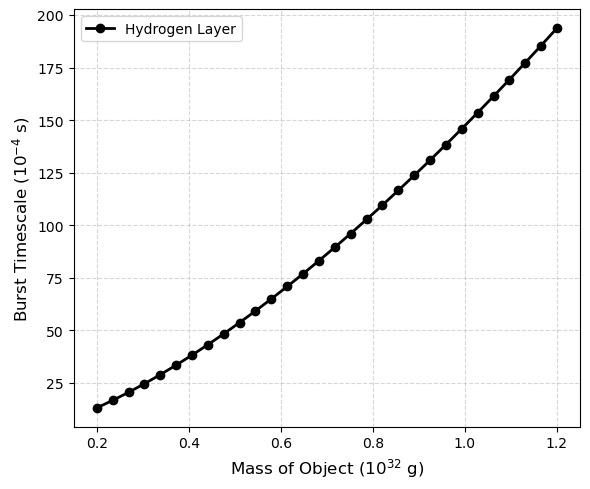}
    \caption{Plot between Mass of Object and Burst Timescale for H layer}
    \label{fig:H layer}
\end{figure}

\begin{figure}[H]
    \centering
    \includegraphics[width=0.6\linewidth]{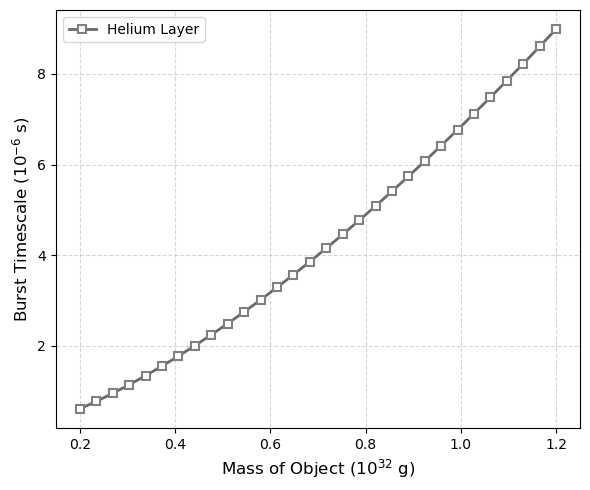}
    \caption{Plot between Mass of Object and Burst Timescale for He layer}
    \label{fig:He layer}
\end{figure}

\begin{figure}[H]
    \centering
    \includegraphics[width=0.6\linewidth]{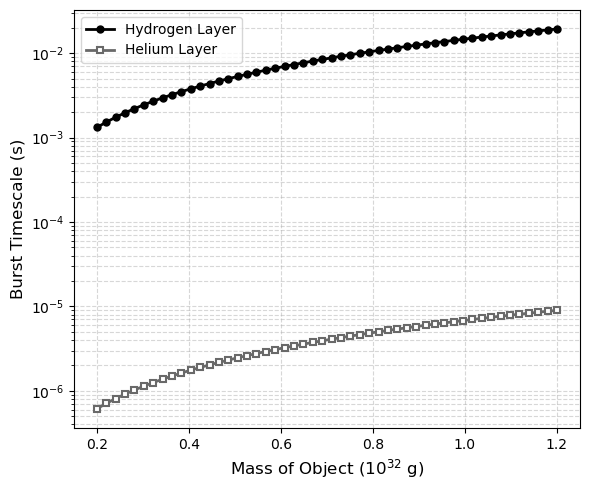}
    \caption{Plot comparing the Hydrogen and Helium Burst Timescales as a function of object mass.}
    \label{fig:Comparison Plot}
\end{figure}

The Galactic Center GeV excess is a surplus of energy near the galactic center in gamma rays. This was initially reported in the energy range of 300 MeV to 10 GeV \cite{300-10GeV}, but subsequent analyses with updated LAT data have placed this excess within the 1--3 GeV range \cite{Daylan_2016, Calore_2015}. Early investigations by Goodenough \textit{et al.} \cite{goodenough2009possibleevidencedarkmatter} and Vitale \cite{early2} showed that the dark matter annihilation process could explain the excess.\\
We will now determine if the typical energies of the radiation for these compact fermion dark matter objects agree with the observed Galactic Center GeV excess.
Using Wien's displacement law based on the estimated temperature, we determine the peak frequency of both layers, which in turn can be used to calculate the energies of these photons for comparison with the Galactic Center GeV excess. The calculated values are presented in Table~\ref{tab:energy_layers}.
\begin{table}[!ht]
\centering
\renewcommand{\arraystretch}{1.3}
\begin{tabular}{|c|c|c|}
\hline
\textbf{Parameter} & \textbf{Hydrogen ($H$)} & \textbf{Helium ($He$)} \\
\hline
Temperature (K) & $1.192 \times 10^{11}$ & $4.770 \times 10^{11}$ \\
\hline
Frequency (Hz) & $6.95 \times 10^{21}$ & $2.78 \times 10^{22}$ \\
\hline
Energy (eV) & $2.88 \times 10^{7}$ & $1.15 \times 10^{8}$ \\
\hline
\end{tabular}
\caption{Peak frequencies and corresponding energies for the hydrogen and helium layers.}
\label{tab:energy_layers}
\end{table}\\
Although the estimated photon energies fall within the gamma-ray regime, they remain significantly below the characteristic $1$--$3$ GeV energy range associated with the Galactic Center excess. Consequently, the compact fermionic dark matter objects considered in the present model are unlikely to account for the observed excess emission.\\
Radiative power from a spherical shell of radius $r$ at temperature $T$ is
\begin{equation}
    P = 4\pi \sigma r^2 T^4
\end{equation}
and the characteristic emission timescale is defined as
\begin{equation}
    \tau = \frac{E_{\rm released}}{P} = \frac{E_{\rm released}}{4\pi \sigma r^2 T^4}
    \label{eq:timescale}
\end{equation}
Using Eq.~\eqref{eq:timescale}, the total energies released and the burst timescales for each layer are computed as shown in Table~\ref{tab:timescale_burst}.
\begin{table}[h!]
    \centering
    \renewcommand{\arraystretch}{1.3}
    \begin{tabular}{|c|c|c|}
    \hline
    \textbf{Parameter}     & \textbf{Hydrogen ($H$)} & \textbf{Helium ($He$)} \\ \hline
    Energy Released (erg)  & $1.236 \times 10^{45}$  & $1.05 \times 10^{47}$  \\ \hline
    Burst Timescale (s)    & $9.7\times 10^{-4}$                    & $4.5\times 10^{-7}$                 \\ \hline
    \end{tabular}
    \caption{Computed energy releases and corresponding burst timescales for each layer.}
    \label{tab:timescale_burst}
\end{table}\\
Estimating the burst timescale is essential for assessing its observability. The calculated burst durations are extremely short, making the direct detection of such events unlikely with present instruments.\\
Reevaluating the constraints imposed by microlensing surveys gives a narrow range of feasible objects, which we then employ to compute the corresponding burst timescales in Fig.~\ref{fig:H layer} and Fig.~\ref{fig:He layer}. The comparative behavior of the hydrogen and helium burst timescales is shown in Fig. ~\ref{fig:Comparison Plot}.


\noindent To facilitate a direct comparison between the hydrogen and helium layers, the burst timescales are shown together in Fig.~\ref{fig:Comparison Plot}. The hydrogen layer exhibits systematically longer burst durations, while both timescales increase with object mass according to the adopted scaling relations. The energy released from our object is ranging only from \textbf{28 MeV} to \textbf{115 MeV}, which falls well below the 1–3 GeV range of the Galactic Center excess. Therefore, the compact object that we’ve constrained here cannot account for the observed GeV signal.\\
In addition to purely fermionic dark matter compact objects, we consider the possibility of admixed compact objects containing both fermionic dark matter and baryonic matter. The baryonic fraction $f_{b}$ is used to describe the ratio of the baryonic mass to the total mass of the object, while the rest of the mass comprises the fermion dark matter.  Since the baryonic content may vary depending on the formation history of the object, representative baryonic fractions ranging from 0.1 to 0.9 are considered to investigate their potential contribution to the Missing Baryon Problem.

\section{The Missing Baryon Problem}

The Cosmic Microwave Background (CMB) measurements by WMAP show that baryonic matter makes up around 4.6\% of the energy density of the Universe \cite{komatsu2009five}, which is in line with the prediction of Big Bang nucleosynthesis \cite{fukugita1998cosmic}. But a complete census of stars, cold interstellar matter, and hot intracluster plasma at low redshift shows only around 70\% of this baryonic matter, leaving the other 30\% of the baryons undetected in the local Universe \cite{Shull_2012}.\\
A comprehensive analysis \cite{Shull_2012} leading to a multiphase baryon census using ultraviolet absorption lines of the Ly$\alpha$ forest, O VI doublet, and broad Ly$\alpha$ absorbers combined with ionization and metallicity corrections to derive mass fractions in each gas phase.  They determined that the photoionized Ly$\alpha$ forest at $T\sim10^4$ K contains $28\%\pm11\%$ of $\Omega_b$, O VI–traced warm–hot gas at $T\sim10^{5.3}$–$10^{5.7}$ K contributes $\sim17\%$, and additional baryons reside in the circumgalactic medium ($\sim5\%$), galaxies (stars + cold gas; $\sim7\%$), and the intracluster medium ($\sim4\%$), summing to $\sim71\%$ of the expected baryons and leaving a residual $29\%\pm13\%$ unaccounted for \cite{Shull_2012}.\\
Cosmological hydrodynamic simulations predict that the missing baryons are shock-heated during structure formation and reside in a warm–hot intergalactic medium (WHIM) at $10^5\!<\!T\!<\!10^7$ K, distributed along the filamentary cosmic web \cite{Cen_2006,Davhydro}.  Early reviews \cite{bregmanmbp} emphasized that, despite detections of O VI–traced gas and broad Ly$\alpha$ absorbers, there remained no conclusive evidence for the hotter ($T>10^6$ K) WHIM phase due to instrumental limitations.  Observations carried out with the help of X-rays \cite{Nicastro_2018mbp} found absorbers of O VII in quasar spectra, which were consistent with the predictions of the WHIM columns and could possibly solve the problem of missing baryons in a large proportion. Other studies made on the circumgalactic medium \cite{werkmbp}, based on the HST/COS–Halos survey, found cool halo gas of $T\sim 10^4$ K to have masses of $\gtrsim 6.5\times10^{10} M_\odot$ around $L^*$ galaxies, implying that galactic halos contain a large amount of baryons, yet cannot resolve the cosmic baryon crisis. Nevertheless, the Missing Baryon Problem persists ($\sim30\%$).\\

\noindent One possible explanation for a fraction of the missing baryons is that they may reside in compact objects composed of both fermionic dark matter and baryonic matter. However, since the present work does not address the physical formation mechanism of such admixed compact objects, this should be regarded only as a phenomenological possibility rather than a demonstrated solution to the missing baryon problem. We consider admixed compact objects containing a baryonic fraction ' $f$ ' of baryonic particles, which is being mixed with a $(1-f)$ fraction of DM particles. Then the radius of such objects is given by \cite{universe9090401},\\
\begin{equation}
    R = \frac{\hbar^2 M_T^{-\frac{1}{3}}}{G} \left( \frac{(1-f)^{\frac{2}{3}}}{m_D^{\frac{8}{3}}} + \frac{f^{\frac{5}{3}}}{(1-f)m_p^{\frac{8}{3}}} \right)
\end{equation}
where $m_D$ is the fermionic DM particle mass, $m_p$ is the BM particle mass, $M_T$ is the total mass of the object given by 
\begin{equation}
    M_T = \frac{M_{pl}^3}{m^2_{eff}}
\end{equation}
and $m_{eff}$ is the effective mass of constituent particles given by
\begin{equation}
    m_{eff} = (1-f)m_D + fm_p 
\end{equation}
With the help of the above equations, and considering the feasible mass of DM particle \textbf{10 GeV}, we calculate the mass and radius of compact objects composed of BM and DM particles were computed for baryonic mass fractions $f_b$ ranging from 0.1 to 0.9. The object mass rises with increasing baryonic fraction $f_b$ and can have maximum value of $\sim10^{33}$g (hundreds of masses of Jupiter) as can be seen from Table~\ref{tab:mbp}.

\begin{figure}[h!]
    \centering
    \includegraphics[width=0.7\linewidth]{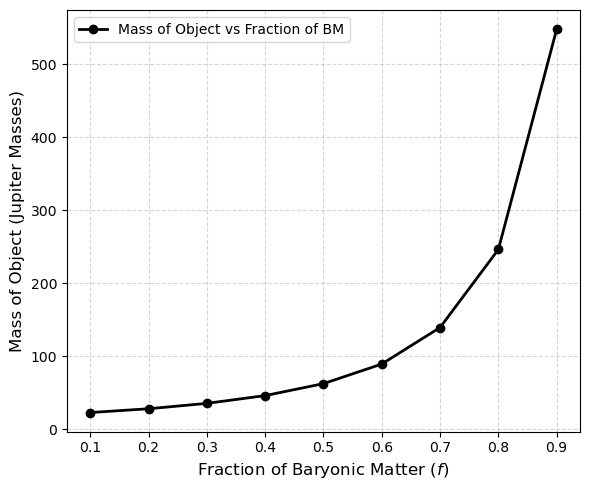}
    \caption{Plot between Mass of the primordial planet and Fraction of baryons}
    \label{fig:fraction}
\end{figure}

\begin{table}[H]
\centering
\renewcommand{\arraystretch}{0.95}
\begin{tabular}{|c|c|c|c|c|}
\hline
Fraction of BM & Fraction of DM & Effective Mass (GeV) & Mass of Object (g) & Radius (cm) \\
\hline
0.1 & 0.9 & 9.09 & $4.28 \times 10^{31}$ & $3.52 \times 10^4$ \\
\hline
0.2 & 0.8 & 8.19 & $5.27 \times 10^{31}$ & $1.11 \times 10^5$ \\
\hline
0.3 & 0.7 & 7.28 & $6.67 \times 10^{31}$ & $2.29 \times 10^5$ \\
\hline
0.4 & 0.6 & 6.38 & $8.69 \times 10^{31}$ & $3.94 \times 10^5$ \\
\hline
0.5 & 0.5 & 5.47 & $1.18 \times 10^{32}$ & $6.17 \times 10^5$ \\
\hline
0.6 & 0.4 & 4.57 & $1.69 \times 10^{32}$ & $9.26 \times 10^5$ \\
\hline
0.7 & 0.3 & 3.66 & $2.64 \times 10^{32}$ & $1.38 \times 10^6$ \\
\hline
0.8 & 0.2 & 2.75 & $4.67 \times 10^{32}$ & $2.13 \times 10^6$ \\
\hline
0.9 & 0.1 & 1.84 & $1.04 \times 10^{33}$ & $3.97 \times 10^6$ \\
\hline
\end{tabular}
\caption{Mass and Radius of the planet made up of baryons and DM particles}
\label{tab:mbp}
\end{table}
\noindent Fig.~\ref{fig:fraction} presents a plot of object mass versus the fraction of BM, illustrating that as the BM fraction increases, the total mass of the object also rises reaching values as high as 500 times the mass of Jupiter.

\noindent Using the masses obtained for different fractions of BM, we evaluate the corresponding microlensing signature for these objects. these results are then used to assess their potential contribution to the missing baryon inventory and are summarized in Table~\ref{tab:microlensing}. We calculate its contribution to the Missing Baryon Problem through the following steps:

\begin{table}[ht]
\centering
\renewcommand{\arraystretch}{1.5}
\begin{tabular}{|c|c|c|c|c|c|}
\hline
BM Fraction & Mass(g)  & Einstein Radius (AU) & Microlensing Time (days) & Events-I & Event-II \\ \hline
0.1 & 4.28$\times 10^{31}$    & 1.448             & 11.40                  & 89.5186      & 156.0846    \\ \hline
0.2 & 5.27$\times 10^{31}$    & 1.609             & 12.66                  & 85.9642      & 140.5307    \\ \hline
0.3 & 6.67$\times 10^{31}$    & 1.809             & 14.24                  & 80.2199      & 124.9769    \\ \hline
0.4 & 8.69$\times 10^{31}$    & 2.066             & 16.26                  & 70.6986      & 109.4230    \\ \hline
0.5 & 1.18$\times 10^{32}$    & 2.408            & 18.95                 & 58.3684      & 93.8691     \\ \hline
0.6 & 1.69$\times 10^{32}$    & 2.887            & 22.72                  & 46.8809      & 78.3152     \\ \hline
0.7 & 2.64$\times 10^{32}$   & 3.602             & 28.35                  & 38.4210      & 62.7613     \\ \hline
0.8 & 4.67$\times 10^{32}$    & 4.789             & 37.69                  & 29.1567      & 47.2075     \\ \hline
0.9 & 1.04$\times 10^{33}$    & 7.143             & 56.22                  & 19.4328      & 31.6536     \\ \hline
\end{tabular}
\caption{Microlensing calculations for different masses. Events-I is obtained by integrating \cite{Croon_2020} and Events-II is obtained through the EROS (optical depth) method \cite{Tisserand_2007} }
\label{tab:microlensing}
\end{table}

\noindent The total Baryonic mass of the universe is $10^{56}g$ and the missing percentage of missing baryons is 29\%. Therefore, the mass of missing baryons is found to be,
$$M_{missing} = 0.29 \times 10^{56} = 2.9 \times 10^{55}g $$
Baryonic mass per object is given as,

$$M_{baryonic} = f_b \times M_{object} $$
and from there we can calculate the number of objects given by,
\begin{equation}
    N_{objects} = \frac{M_{missing}}{M_{object}}
\end{equation}
To estimate the potential contribution of these admixed compact objects to the missing baryon inventory, we define the fractional contribution as

$$ Percentage = \frac{N_{objects} \times M_{baryonic} }{M_{missing}} \times 100     $$
The calculated values are tabulated in Table{~\ref{table:baryonic-contribution}.

\begin{table}[H]
    \centering
\renewcommand{\arraystretch}{1}
\begin{tabular}{|c|c|c|c|c|}
\hline
$f_b$ & $M_{\text{obj}}$ (g) & $N_{\text{objects}}$ & $M_\text{baryonic}$ (g) & Percentage Contribution\\ \hline
0.1 & $4.28 \times 10^{31}$ & $6.78 \times 10^{23}$ & $4.28 \times 10^{30}$ & $10\%$ \\ \hline
0.2 & $5.27 \times 10^{31}$ & $5.50 \times 10^{23}$ & $1.05 \times 10^{31}$ & $20\%$ \\ \hline
0.3 & $6.67 \times 10^{31}$ & $4.35 \times 10^{23}$ & $2.00 \times 10^{31}$ & $30\%$ \\ \hline
0.4 & $8.69 \times 10^{31}$ & $3.34 \times 10^{23}$ & $3.48 \times 10^{31}$ & $40\%$ \\ \hline
0.5 & $1.18 \times 10^{32}$ & $2.46 \times 10^{23}$ & $5.90 \times 10^{31}$ & $50\%$ \\ \hline
0.6 & $1.69 \times 10^{32}$ & $1.72 \times 10^{23}$ & $1.01 \times 10^{32}$ & $60\%$ \\ \hline
0.7 & $2.64 \times 10^{32}$ & $1.09 \times 10^{23}$ & $1.85 \times 10^{32}$ & $70\%$ \\ \hline
0.8 & $4.67 \times 10^{32}$ & $6.21 \times 10^{22}$ & $3.74 \times 10^{32}$ & $80\%$ \\ \hline
0.9 & $1.04 \times 10^{33}$ & $2.79 \times 10^{22}$ & $9.36 \times 10^{32}$ & $90\%$ \\ \hline
\end{tabular}
\caption{Contribution to the Missing Baryons by the objects}
\label{table:baryonic-contribution}
\end{table}

\begin{figure}[!ht]
    \centering
    \includegraphics[width=0.6\linewidth]{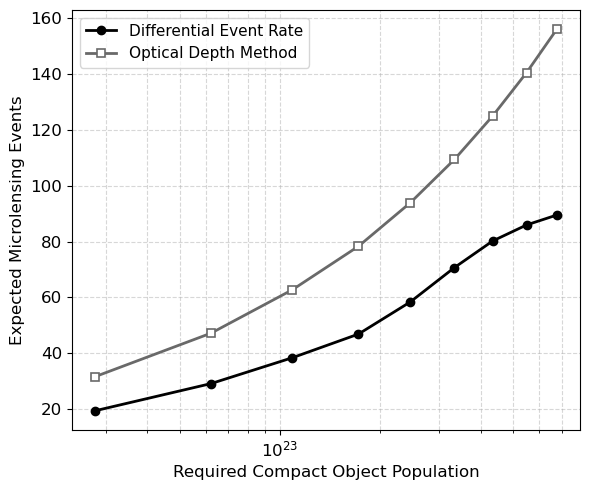}
    \caption{Expected microlensing events as a function of the required number of admixed compact objects for different baryonic fractions ($f_b=0.1$--$0.9$). The event rates are computed independently using the differential event-rate formalism and the optical-depth method towards the Large Magellanic Cloud (LMC).}
    \label{fig:Population vs Events}
\end{figure}

\noindent However, as the baryonic fraction increases, the mass of the admixed compact objects increases and hence fewer such compact objects are needed to explain the baryonic mass. This is clearly illustrated in Fig.~\ref{fig:Population vs Events}, where we observe that the compact object population decreases as the baryonic fraction increases. The above statement must be taken in context that while the admixed compact objects can form in the discrete baryonic fractions, the compact objects can actually form in any other baryonic fraction (say, 35\%, 64\%), etc.

\section{Conclusions}

We have investigated the hypothesis that fermions may assemble into compact self gravitating objects supported by degeneracy pressure. \\
From our analysis of gravitational microlensing signatures, we identify the most feasible mass for compact dark matter objects to be $M_{DM object} = 1.63 \times 10^{-2} M_{\odot} $
which corresponds to a fermionic  dark matter particle mass of 10 GeV. The lower-mass candidates ($10^{-4} M_{\odot}$ or $10^{-6} M_{\odot}$) are excluded due to their predicted overproduction of microlensing events.\\
To further characterize these objects, we modeled their structural and thermodynamic evolution, including mass accretion from surrounding baryonic matter and the resulting energy release. Our calculations reveal that although these compact dark matter objects can accumulate significant layers of hydrogen and helium, the peak energies emitted from these layers are limited to the tens to low hundreds of MeV range. This emission spectrum lies substantially below the 1-3 GeV energy band associated with the Galactic Center GeV Excess. Thus, we have found that the energy released by fermion based objects are much significantly lower than the observed GeV excess, hence our objects cannot account for this discrepancy.\\
With respect to a phenomenological mixture of dark matter and baryonic matter in such admixed compact objects, our calculations show that these objects could host a considerable amount of the universe’s missing baryonic matter. Provided such admixed compact objects can form via a physical process, these objects could serve as a possible place for the missing baryonic matter. The current work has not investigated the processes of formation of such objects; therefore, this result should be treated only as a phenomenological implication and not a solution to the missing baryon problem.

\bibliographystyle{ieeetr}
\bibliography{mybib}

\end{document}